\documentclass[12pt]{amsart}
\usepackage{amssymb,righttag}
\font\smc=cmcsc10 at 12pt
\font\eightsmc=cmcsc8 at 12pt
\font\bigit=cmti10 at 12pt
\newtheorem{thm}{Theorem}[section]

\newtheorem{defin}[thm]{Definition}

\theoremstyle{remark}
\newtheorem{remark}[thm]{Remark}
\let\cal=\mathcal
\def\R{{\Bbb R}}\def\Z{{\Bbb Z}}\def\C{{\Bbb C}}
   \def\im{\Im m\,}
\def\cL{{\cal L}}\def\cU{{\cal U}}\def\cE{{\cal E}}\def\cF{{\cal F}}
\def\cO{{\cal O}}\def\cS{{\cal S}}

\def\Cal#1{{\cal #1}}
\def\fF{{\frak  F}} \def\fC{{\frak  C}}
\let\o=\operatorname
\def\oskip{\par\addvspace{4mm}\par}
\def\doublecheck#1{{\kern2.1pt\hbox{$\check{\kern-2.1pt\hbox{$\check #1$}}$}}}
\def\X{{\widehat X}}
\def\bysame{$\raise.2em\hbox to 3em{\hrulefill}$\thinspace, }
\begin{document}
\pagestyle{plain}\footnotesep=12pt
\vbox to1cm{\vfill}
\begin{center}
{\bf CATEGORIAL MIRROR SYMMETRY}
\par\medskip
{\bf FOR K3 SURFACES} \par\bigskip\bigskip
\par\bigskip\bigskip
{\smc C. Bartocci,\ddag}\ {\smc U. Bruzzo,\S\ddag}\ \ and \ {\smc G.
Sanguinetti\P}
\par\bigskip
{\ddag\thinspace Dipartimento di Matematica, Universit\`a degli Studi}
\par{di Genova, Via Dodecaneso 35, 16146 Genova, Italy}
\par\smallskip
{\S\thinspace Scuola Internazionale Superiore di Studi Avanzati }
\par {(SISSA), Via Beirut 2-4, 34014 Trieste, Italy}
\par\smallskip
{\P\thinspace Mathematical Institute, University of Oxford,}\par
{24-29 St.~Giles', Oxford OX1 3LB, U.K.}
\par\smallskip
{E-mail addresses: {\tt bartocci@dima.unige.it}, {\tt bruzzo@sissa.it},
{\tt sanguine@maths.ox.ac.uk}}
\end{center}
\par\oskip\oskip
{\footnotesize\baselineskip=11.5pt\par\leftskip=40pt\rightskip=40pt
{\eightsmc Abstract.} We
study the structure of a modified  Fukaya category $\fF(X)$ associated with
a  K3
surface $X$, and prove that whenever $X$ is an elliptic K3 surface with a
section, the derived category of $\fF(X)$ is equivalent to a  subcategory
of the derived
category ${\bold D}(\X)$ of coherent sheaves on the mirror K3 surface
$\X$.\par}
\oskip\oskip
\section{Introduction}
In 1994  M. Kontsevich conjectured that a proper mathematical  formulation of
the mirror conjecture is provided by an equivalence between Fukaya's
category of a
Calabi-Yau  manifold $X$ and the derived category of coherent sheaves of the
mirror Calabi-Yau manifold $\X$ \cite{K}.  Thus in some sense  mirror symmetry
relates  the symplectic structure of a Calabi-Yau manifold with the holomorphic
structure of its mirror. It is expected that special Lagrangian tori on $X$
are mapped by mirror symmetry to skyscraper sheaves on the mirror $\X$.

This conjecture found some physical evidence with the discovery of D-branes
and the description of their role in mirror symmetry \cite{OOY,SYZ}.
Moreover, in a recent paper \cite{PZ}  Kontsevich's conjecture has been proved
in the case of the simplest  Calabi-Yau manifolds, the elliptic
curves.

Our approach to mirror symmetry follows the geometric interpretation
due to Strominger, Yau and Zaslow \cite{SYZ}. According to their construction,
given a Calabi-Yau manifold admitting a foliation in special Lagrangian tori,
its mirror manifold should be obtained by relative T-duality. In the case
of K3 surfaces this formulation has been given a rigorous treatment in
\cite{BBHM,BS}, proving that Strominger, Yau and Zaslow's approach is
consistent
with previous descriptions of mirror symmetry \cite{D} (this is also related
to work by Aspinwall and Donagi \cite{AD}).

We show here how the constructions described in \cite{BBHM,BS} can be given a
categorial interpretation which provides a proof of Kontsevich's conjecture in
the case of K3 surfaces. More precisely, we show that, under some
assumptions which will be spelled out in the following Sections,  the
derived category of a Fukaya-type category  built out of special Lagrangian
submanifolds of an elliptic K3 surface $X$ is equivalent to a subcategory
of the
derived category of coherent sheaves on the mirror surface $\X$.
This subcategory is formed by the complexes of sheaves whose zeroth Chern
character vanishes.

\oskip
\section{Special Lagrangian submanifolds and Fukaya's category}
\begin{defin} Let $X$ be a Calabi-Yau $n$-fold, with K\"ahler form $\omega$ and
holomorphic $n$-form $\Omega$. A (real) $n$-dimensional submanifold
$\iota\colon M\hookrightarrow X$
of $X$ is said to be {\it special Lagrangian} if the following two conditions
are  met:

---  $X$ is Lagrangian in the symplectic structure given by $\omega$,
i.e.  $\iota^\ast
\omega =0$;

--- there exists a  multiple $\Omega '$ of $\Omega$ such that
$\iota^\ast\im \Omega '=0$.
\end{defin}
It can be shown  that the second condition is equivalent  to requiring that
the real part of $\Omega '$ restricts to the volume form of $M$ induced by
the
Riemannian metric of $X$. This exhibits special Lagrangian submanifolds
as a special type of calibrated submanifolds \cite{HL}.

There are not many explicit examples of  special Lagrangian submanifolds.  The
simplest ones are the 1-dimensional submanifolds of an elliptic curve:
the first condition is trivial, and the  multiple $\Omega'$ of the global
holomorphic one-form $\Omega$ is readily obtained by a holomorphic change of
coordinates in the universal covering of the elliptic curve.
Additional examples are provided by Calabi-Yau manifolds equipped with an
antiholomorphic  involution. Since the involution changes the sign of both
the K\"ahler form and   the imaginary part of the holomorphic $n$-form,  the
fixed point sets of the involution are special Lagrangian submanifolds.
A third example, and the most relevant in our case,  arises
when considering Calabi-Yau manifolds  endowed with a hyper-K\"ahler
structure. This is
always the case in dimension 2, i.e.~for K3 surfaces. In this case special
Lagrangian submanifolds are just holomorphic submanifolds with respect to a
different complex structure compatible with the same hyper-K\"ahler metric.
This example will
be discussed at length in the next section.

Special Lagrangian submanifolds have received   remarkable attention in physics
since the  appearance of D-branes in string theory, and
especially since their role turned out to be of a primary importance for
the mirror
conjecture \cite{BBS,SYZ}. D-branes are special Lagrangian
submanifolds of the Calabi-Yau manifold which serves as compactification space,
and are equipped with  a flat $U(1)$ line bundle. In the physicists'
language, special
Lagrangian submanifolds of the compactification space are
associated with physical states which retain  part of the supersymmetry of
the vacuum. For this (and other related) reasons, special Lagrangian
submanifolds
are often called {\it supersymmetric cycles}, or also {\it BPS states}.

Fukaya's category, whose objects are Lagrangian submanifolds of a
symplectic manifold,
was introduced in connection with Floer's homology  \cite{F}. Here,
following  closely the exposition of \cite{PZ}, we
offer  a description of a modified Fukaya  category, built out of the special
Lagrangian submanifolds of a
Calabi-Yau manifold $X$. We shall call this the {\it special Lagrangian Fukaya
category} (SLF category for short) of $X$, and will denote it by $\fF(X)$.
The objects in
$\fF(X)$ are pairs ($\cL ,\cE )$, where
$\cL$ is a special Lagrangian submanifold of
$X$, and
$\cE$ is a flat vector bundle on $\cL$. The morphisms in this category are
a little bit
more complicate to define. Since special Lagrangian submanifolds are
$n$-cycles in a
compact complex $n$-dimensional manifold, two special Lagrangian cycles
generically intersect at a
finite number of points. The basic concept is that a morphism between two
objects in
the SLF category is a way to pass from the vector bundle defined on one
cycle to the
bundle on the other.

\begin{defin}Let $\cU _1=(\cL _1, \cE _1)$, $\cU _2=(\cL _2, \cE _2)$ be
two objects in
the SLF category. Then the space of morphisms $\o{Hom}(\cU _1, \cU _2)$ is
defined to
be
$$\o{Hom}(\cU _1, \cU _2)=
\oplus _{x\in \cL _1\cap \cL _2} \o{Hom}(\cE _1|_x, \cE _2|_x).$$
\end{defin}
Thus the space of morphisms between two objects in the SLF category turns
out to be
the  direct sum of vector spaces, each one being the space of homomorphisms
between the
fibers of the two vector bundles at the intersection points of the two
special Lagrangian cycles.

{\bigit Maslov index.} The space of morphisms between two objects  is
naturally
graded over $\Z$  by the  Maslov index of the tangent spaces to the special
Lagrangian submanifolds at  the intersection points  \cite{PZ}. Let us
recall some basic
facts about the Maslov index. Let $V$ be a $2n$-dimensional real symplectic
vector space, and
denote by
${\cal G}(V)$ the Grassmannian  of Lagrangian $n$-planes in $V$. One has an
isomorphism
${\cal G}(V)\simeq U(n)/O(n)$, so that $\pi_1({\cal G}(V))\simeq\Z$. The
Maslov index
is the unique integer-valued function on the space of loops in ${\cal
G}(V)$   satifying
some naturality conditions \cite{MDS} which include its homotopic invariance;
thus the Maslov index  provides
an explicit isomorphism  $\pi_1({\cal G}(V))\to\Z$. In order to define a
Maslov index
for the intersection of Lagrangian cycles one has to  slightly modify
its definition so as to consider  open paths. One first   notices that
the Lagrangian Grassmannian is naturally stratified  by the dimension of the
intersection of the Lagrangian $n$-planes with a fixed Lagrangian
$n$-plane. Then one
can define
a Maslov index for the intersection of two Lagrangian planes  as a $\Z$-valued
function one the space of paths in ${\cal G}(V)$ which is homotopy invariant
under deformations of the paths that do not move the extrema out of their
strata.

(Actually one should consider a Grassmannian of special Lagrangian (rather
than just Lagrangian)
planes, and restrict the Maslov index to it. This will be done in the next
section in the case
of K3 surfaces.)

{\bigit $A^\infty$ structure.} Strictly speaking an SLF category, as it
happens with
 ordinary Fukaya categories,  is not a   category at all, since in
general   the
composition of morphisms  fails to be associative. Associativity is
replaced by a more
complicated property, which makes  Fukaya's ``category'' into an $A^\infty$
category.
\begin{defin} An $A^{\infty}$ category $\fF$ consists of

 a class of objects $Ob(\fF)$;

  for any two objects $\Cal X,\Cal Y$, a $\Z$-graded abelian group of
morphisms 
$\o{Hom} (\Cal X,\Cal Y)$;

  composition maps
$$m_k:\o{Hom}(\Cal X_1, \Cal X_2)\otimes\dots \otimes\o{Hom}(\Cal X_k, \Cal
X_{k+1})\to
\o{Hom}(\Cal X_1, \Cal X_{k+1}),,\quad k\ge 1,$$
of degree $2-k$, satisfying the condition
\begin{align}
\sum_{r=1\dots n\atop s=1\dots n-r+1}(-1)^{\epsilon}\, m_{n-r+1} &
(a_1\otimes\dots
\otimes a_{s-1}\otimes m_r(a_s\otimes\dots \nonumber \\ & \dots
 \otimes a_{s+r-1})\otimes a_{s+r}\otimes\dots \otimes a_n) =0
\label{e:genass}\end{align}
for all $n\ge 1$, where
$$\epsilon =(r+1)s+r(n+\sum_{j=1}^{s-1}\o{deg}(a_j))\,.$$ \end{defin}
Condition (\ref{e:genass}) implies that $m_1$ is a
coboundary operator. The vanishing of the morphism $m_1$, together with
condition (\ref{e:genass}) for the morphism $m_3$, implies that the composition
law given by $m_2$ is associative.

Let us see how this $A^{\infty}$ structure arises in Fukaya's category. Let
us assume that
 the first object $\Cal X_1$
and the last object
$\Cal X_{k+1}$ have a nonvoid intersection, otherwise $\o{Hom}(\Cal X_1,
\Cal X_{k+1})=0$ and
the composition map is trivial. The composition maps
are explicitly described as follows: Let
$u_j=(a_j, t_j)\in \o{Hom}(\cU _j, \cU_{j+1})$, where $a_j\in \cL _j\cap
\cL _{j+1}$
and
$t_j\in \o{Hom}(\cE _j|_{a_j}, \cE _{j+1}|_{a_j})$. One defines
\begin{equation*}
m_k(u_1\otimes \dots \otimes u_k)=\sum _{a_{k+1}\in \cL _1\cap \cL
_{k+1}}(C(u_1,\dots ,
u_k), a_{k+1}). \end{equation*}
Here one  has
\begin{equation*}
C(u_1,...,u_k,a_{k+1})=\sum_{\phi}\pm \exp[2\pi i(\int \phi ^*\omega ^c)]
P\exp[\oint \phi^*
\beta].
\end{equation*}
This requires some explanation. The sum is performed over holomorphic and
antiholomorphic maps $\phi$ from the disc $D^2$ into the manifold $X$, up to
projective
equivalence, with the following boundary condition: there are $k+1$ points
$p_j=
e^{2\pi
\alpha_j}\in S^1= \partial D^2$ such that
$\phi(p_j)=a_j$ and $\phi(e^{2\pi
\alpha})\in \cL_j$ for $\alpha\in (\alpha _{j-1},\alpha _j).$
The two-form $\omega^c$ appearing in (2) is the complexified K\"ahler form,
while $\beta
$ is the connection of the bundle restricted to the image of the boundary
of the disc.
$P$ represents a path-ordered integration, defined by
\begin{align*} & P\exp({\oint \phi^*
\beta})\\ & =\exp({\int_{\alpha _k}^{\alpha _{k+1}}\beta _k d\alpha})\,
t_k\,\exp({\int_{\alpha _{k-1}}^{\alpha _{k}}\beta _{k-1}
d\alpha})\,t_{k-1}\,...t_1\,
\exp({\int_{\alpha _{k+1}}^{\alpha _{1}}\beta _1 d\alpha})\,.\end{align*}

\oskip\section{The special Lagrangian Fukaya  category for K3 surfaces}
The main purpose
of this section is to give a description of the SLF  category when
the Calabi-Yau manifold is a K3 surface $X$. In this case,   due to the
fact that K3
surfaces admit   hyper-K\"ahler metrics, special Lagrangian submanifolds are
very easily
exhibited.  Let us denote by $\omega$ the K\"ahler form associated with   given
hyper-K\"ahler metric and complex structure. One also has a holomorphic 2-form
$\Omega=x+iy$. The three
elements $\omega, x, y$ can be regarded as vectors in the cohomology space
$H^2(X,\R)$; if the latter is equipped with the scalar product of signature
(3,19) induced
by the intersection form on $H^2(X,\Z)$, these three elements are
spacelike, and generate
a 2-sphere which can be identified with the set
of complex structures  compatible with the
fixed hyper-K\"ahler metric.

It is very easy to check that  what is special Lagrangian
in the original complex structure is holomorphic in the complex structure
in which the
roles of
$\omega$ and $x$ are exchanged (up to a sign) \cite{HL} (this corresponds
to a rotation of
$90$ degrees around the $y$ axis). We shall call such a change of complex
structure a
{\it hyper-K\"ahler rotation.}

We want in particular to consider {\it elliptic} K3 surfaces $X$ which admit a
section.\footnote{This means that there exists an epimorphism $p\colon
X\to{\Bbb P}^1$
whose generic fiber is a smooth elliptic curve and admitting a section
$e\colon{\Bbb
P}^1\to X$.} K3 surfaces arising as compactification spaces of string
theories which
admit mirror partners are always of this type   \cite{SYZ}.
So let us consider a K3 surface $X$ that in a complex structure $I$ is elliptic
and has a section. Let us denote by $X_I$ this K3 surface. The Picard group
of $X_I$ is generated by the section, by
the divisor of the generic fiber, and by the irreducible components of the
singular
fibers that
do not intersect the section.\footnote{Actually one may have further
generators of the
Picard group provided by additional sections of the projection $p\colon
X\to{\Bbb P}^1$.}
If we perform the hyper-K\"ahler rotation described above, and call $J$ the
new complex
structure, the submanifolds which were holomorphic in the complex structure
$I$ are now
special  Lagrangian. Assuming that $X_J$ is elliptic as well, it has
been shown \cite{BS} that this hyper-K\"ahler rotation reproduces, at the
level of the
Picard lattice of an elliptic K3 surface, the effects of mirror symmetry
previously
described in an algebraic way \cite{D}. So the varieties $X_I$ and $X_J$
can be
regarded as a {\it mirror pair} of K3 surfaces.

In this way one has a very precise picture of the configuration of  special
Lagrangian
submanifolds of $X_J$. Moreover, the flat vector bundles one considers
on special Lagrangian submanifolds  of $X_J$ are (flat) holomorphic bundles
in the
complex structure $I$.

On a K3 surface  the $A^\infty$ structure of  the SLF category
turns out to be trivial, that is,  the SLF category is a true category. In
fact due to
the hyper-K\"ahler structure of a K3 surface $X$, the Grassmaniann of
special Lagrangian
subspaces of the tangent space to $X$ at a point reduces to a copy of
${\Bbb P}^1$,
hence   is simply connected. Moreover, special Lagrangian 2-cycles always
intersect
transversally, so there is no stratification, and the Maslov index is
trivial
(cf.~\cite{K2}).
The $\o{Hom}$ groups in the SLF category have trivial grading, so $m_k=0$ for
$k\ne 2$, while
condition (\ref{e:genass}) for $m_3$ yields the associativity of the
composition of morphisms.

The triviality of this Fukaya category for K3 surfaces may be related, via
Sadov's claim \cite{Sad} that the Floer homology of an almost K\"ahler
manifold $X$ with coefficients in the Novikov ring of $X$ is equivalent to the
quantum cohomology of $X$, to the triviality of the quantum cohomology of K3
surfaces.

\oskip\section{The special Lagrangian Fukaya category and the derived
category of coherent sheaves}
We want  now to describe a construction  which exhibits
the relationship between the SLF category of a K3 surface and the derived
category of coherent sheaves on the mirror K3 surface.

We start by briefly recalling the definition of derived category of an
abelian category
$\frak A$ (cf.~\cite{V}). One starts from the category
${\bold K}({\frak A)}$ whose objects are complexes of objects in $\frak A$,
while the
morphisms are morphisms of complexes identified up to homotopies. Let ${\bold
{Ac}}({\frak A)}$ be the full subcategory of
${\bold K}({\frak A)}$ formed by acyclic complexes (i.e.~complexes such that
all cohomology objects vanishes). The derived category ${\bold D}({\frak A)}$
is by definition the quotient ${\bold K}({\frak A)}/{\bold {Ac}}({\frak A)}$.
A morphism between two objects $[{\cal X}]$, $[{\cal Y}]$ in ${\bold D}
({\frak A)}$ is represented by a diagram of morphisms in ${\bold K}({\frak A)}$
$$ {\cal X} \stackrel{q}{\longleftarrow}  {\cal Z}
\stackrel{m}{\longrightarrow}   {\cal Y} $$
where $q$ is a quasi-isomorphism, i.e., a morphism which induces an
isomorphism between the cohomology objects of ${\cal X}$ and ${\cal Y}$. Two
objects ${\cal X}$, ${\cal Y}$ in ${\bold K}({\frak A)}$ turn out to be
equivalent in ${\bold D}({\frak A)}$ whenever they are quasi-isomorphic, that
is, whenever there is a diagram as above where $m$ is also a
quasi-isomorphism. If there exists a quasi-isomorphism between two
complexes, these represent isomorphic objects in ${\bold D}({\frak A)}$.

Now we consider a K3 surface $X$ with a fixed hyper-K\"ahler metric, and
a compatible complex structure $J$. If we start from an object $(\cL,\cE)$
in the SLF
category $\fF(X_J)$, where $\cL$ is a special Lagrangian submanifold of
real dimension
2, and $\cE$ a flat rank $n$ vector bundle on $\cL$, in the complex structure
$I$ obtained by performing a hyper-K\"ahler rotation $\cL$ is a divisor, and
$\cE$
may be regarded  as a coherent sheaf on $X_J$ concentrated on $\cL$, whose
restriction to $\cL$ is a rank $n$ locally free sheaf.
This operation is clearly functorial: the sheaf of homomorphisms between
two such
objects is a torsion sheaf concentrated on the points where the
two divisors intersect. The stalks at such points are precisely the
homomorphisms
between the stalks of the two coherent sheaves.
Thus the hyper-K\"ahler rotation induces
a functor between the SLF category $\fF(X_J)$ and the category
$\fC(X_I)$ of
coherent sheaves supported on a divisor of $X_I$, whose restriction to the
divisor  is locally free. This functor is clearly faithful, free and
representative
and hence gives an equivalence of the two categories.

\begin{remark} \rm To take  account of the singular divisors in $X$
we should consider {\it torsion-free} sheaves rather than just locally free
ones.
However, since any coherent sheaf on a singular curve over $\C$ has a
projective resolution by locally free sheaves, what we miss by restricting to
locally free sheaves will be recovered when we go to the derived categories.
\qed\label{imp}\end{remark}

The category $\fC(X_I)$ that we obtained via a hyper-K\"ahler rotation
is not abelian (kernels and cokernels of morphisms
do not necessarily lie in the category). In order to introduce a related
derived category, one should find a somehow natural abelian category
$\tilde\fC(X_I)$ containing
$\fC(X_I)$.  The most obvious choice  is the subcategory  of the category
${\frak Coh}(X_I)$ of coherent sheaves on $X_I$ whose objects are sheaves
of rank
0 (in particular we are adding all the skyscraper sheaves).

We assume
that the K3 surface $X_I$ is elliptic and has a section.
Since $X_I$ is elliptic any point $p\in X$
lies on a divisor $D$. The complex
$0\to k_p\to 0$  concentrated  in degree zero, where $k_p$ is the length
one skyscraper at $p$, is
quasi-isomorphic to the complex of sheaves in $\fC(X_I)$
$$0\to \cO_D(-p)\to \cO_D\to 0$$
where $\cO_D$ is the term of degree zero.

Since every coherent sheaf on a smooth curve is the direct sum of a locally
free
sheaf and a
skyscraper sheaf, we  obtain that all coherent sheaves whose
support lies on a divisor are objects of $\tilde\fC(X_I)$.

It is not always true the derived category of an abelian subcategory $\fC'$
of an abelian category $\fC$ is also a subcategory of the derived category of
$\fC$. However, this is indeed the case for the category $\tilde\fC(X_I)$,
as we shall next show.
Let us recall the definition of thick subcategory (cf.~e.g.~\cite{KS}).
\begin{defin} A subcategory
$\fC'$ of a category $\fC$ is said to be thick if for any exact sequence
$\Cal Y\to\Cal Y'\to\Cal W \to\Cal Z\to\Cal Z'$ in  $\fC$ with $\Cal
Y,\Cal Y',\Cal Z,\Cal
Z'$ in $\fC'$  then  $\cal W$ belongs to $\fC'$ as well.
\end{defin}
Now, $\tilde\fC(X_I)$ is a thick subcategory of ${\frak Coh} (X_I)$: in
fact, the
generic stalk of a
sheaf  in $\tilde\fC(X_I)$ is 0, and, since a sequence of sheaves is exact
when it is so at the stalks, this implies that also the generic stalk of
$\Cal W$ is 0,
i.e. $\Cal W$ also is
a rank 0 sheaf. Moreover, $\tilde\fC(X_I)$ is a full subcategory, so that
we can apply the
following theorem
\cite{KS}.
\begin{thm}Let $\fC$ be an abelian category, $\fC'$ a thick full abelian
subcategory. Assume
that for any monomorphism $f\colon \Cal W'\to \Cal W$ with $\Cal W'\in
\operatorname{Ob}(\fC')$, there exists a
morphism $g\colon \Cal W\to \Cal Y$, with $\Cal Y\in
\operatorname{Ob}(\fC')$, such that $g\circ f$ is a
monomorphism. Then the derived category ${\bold D}(\fC')$ is equivalent to the
subcategory of ${\bold D}(\fC)$
consisting of complexes whose cohomology objects belong to $\fC'$. \end{thm}
In our case the
condition of this theorem is easily met, just take for $g$ the evaluation
morphism.
Thus  the derived category built up from $\tilde\fC(X_I)$ is a subcategory
of the
derived category of coherent sheaves.

The image of the
category $\tilde\fC(X_I)$ in cohomology is $H^{1,1}(\Z)\oplus H^4(\Z)$ and
is an
ideal in the algebraic cohomology ring. It is a trivial observation that adding
the unit to an ideal yields the whole ring. Hence, since the Chern map is a
ring
morphism between K-theory and algebraic cohomology, it follows immediately that
by adding the structure sheaf  to $\tilde\fC(X_I)$
we recover the  whole derived category of coherent sheaves.

Adding the structure sheaf of the surface has no motivation from a strictly
geometric viewpoint, but has  physical grounds in the necessity of having
0-branes in the spectrum of the theory. (The association between coherent
sheaves and branes is usually done by taking the Poincar\'e dual of the
support of the coherent sheaf.)

Let us check explicitly that every complex $0\to \cF\to 0$,
where $\cF$ is a coherent sheaf on $X_I$, is quasi-isomorphic to a complex
$$0\to \oplus
\cO_{X_I}\to \cS\to 0\,,$$ where $\cS$ is a coherent sheaf supported on a
divisor.  Let us fix a very ample divisor $H$ in $X_I$. Every coherent sheaf
$\cF$ admits a finite projective resolution by sheaves of the form
$\oplus_{j=1}^r \cO_{X_I}(-m_jH)$ (cf.~\cite{H}).
Moreover, due to the exactness of the sequence $$ 0\to \cO_{X_I}(-m_iH)\to
\cO_{X_I}\to \cO_{m_i H}\to 0\,, $$  the sheaf $\oplus_{j=1}^r
\cO_{X_I}(-m_jH)$ is quasi-isomorphic to a complex $$0\to \oplus
\cO_{X_I}\to \cS\to 0$$
where $\cS$ is a coherent sheaf supported on a divisor  (here $\oplus
\cO_{X_I}$ is concentrated in degree $0$). This proves that the whole derived
category of coherent sheaves is  obtained by complexes whose elements are
either direct sums of the structure sheaf or lie in the image of the SLF
category.

Collecting all these results, we have eventually proved the following fact:
{\it the derived
category of a ``natural abelianization'' of the SLF category
$\fF(X_J)$ is equivalent to a subcategory of the derived category ${\bold
D}(X_I)$ of coherent
sheaves on $X_I$.}

\oskip
\section{Conclusions}
Mirror symmetry yields definite predictions about the transformations of
branes \cite{OOY}, which
can be given a precise mathematical interpretation in terms of
transformations of the
derived category of coherent sheaves. In \cite{BBHM} it was indeed proved
that the action of a
Fourier-Mukai transform on the derived category of coherent sheaves mimics
precisely the
action of mirror symmetry on branes.
In particular, this shows that on an elliptic K3 surface genus 1 special
Lagrangian cycles are mapped to points, which is exactly the
behaviour one  expects from mirror symmetry \cite{M}.

Moreover, one can argue that the very essence of mirror symmetry is an
equivalence between a suitable (derived) version of the Fukaya  category of a
Calabi-Yau manifold $X$  and the derived category of coherent sheaves of the
mirror manifold $\X$.  This is exactly what we have proved when
$X$ is an elliptic K3 surface with a section, admitting also a fibration in
special Lagrangian tori. After performing a hyper-K\"ahler rotation, we map
the SLF category into a category whose ``natural abelianization'' is a thick
full subcategory of the category of coherent sheaves. Now, if we consider an
extension of this category adding the structure sheaf (which seems in some
sense very natural) and derive this, we obtain the whole derived category of
coherent sheaves. Applying a Fourier-Mukai transform (which at the level of
derived categories is an equivalence) we obtain the desired transformation
mapping 2-cycles  of genus 1  to points. If, instead, we do not extend the SLF
category by adding the structure sheaf, we obtain a subcategory of the derived
category of coherent sheaves. This will be mapped by Fourier-Mukai transform
to another subcategory, but again this will show the desired feature of
mapping 2-cycles of genus 1  to points.

\par\smallskip {\bf Acknowledgements.} We thank B.~Dubrovin for valuable
discussions and D.~Hern\'andez Ruip\'erez for his enlightening suggestions.
This research was partly supported
by the research project `Geometria delle variet\`a differenziabili'. The
second author
wishes to thank the School of Mathematical and Computing
Sciences of the Victoria University of Wellington, New Zealand, for the warm
hospitality during the completion of this paper while he was supported
by the Marsden Fund research grant VUW-703.

\oskip


\begin{thebibliography}{22}

\bibitem{AD} Aspinwall, P., and Donagi, R., {\it The heterotic string, the
tangent bundle, and derived categories,} {\tt hep-th/9806094.}

\bibitem{BBHM} Bartocci, C., Bruzzo, U., Hern\'andez Ruip\'erez, D., and
Mu\~noz Porras, J.M., {\it Mirror symmetry on K3 surfaces via Fourier-Mukai
transform,} Commun.~Math.~Phys. {\bf 195} (1998), 79--93; {\tt
alg-geom/9704023.}

\bibitem{BBS} Becker, K., Becker, M., and Strominger, A., {\it Fivebranes,
membranes and non-perturbative string theory,} Nucl.~Phys. {\bf B456} (1995),
130-152; {\tt hep-th/9507158}.

\bibitem{BS} Bruzzo, U., and Sanguinetti, G., {\it Mirror symmetry on  K3
surfaces as a hyper-K\"ahler rotation,} Lett.~Math.~Phys. {\bf 45} (1998),
295--301;
{\tt physics/9802044.}

\bibitem{D} Dolgachev, I.V., {\it Mirror symmetry for  lattice polarized K3
surfaces,} J.~Math.~Sci. {\bf 81} (1996), 2599--2630; {\tt alg-geom/9502005.}

\bibitem{F} Fukaya, K., {\it Morse homotopy, $A^{\infty}$-category and
Floer homologies,} Proceedings of the 1993 GARC Workshop on Geometry and
Topology, Seoul National University.

\bibitem{H} Hartshorne, R., {\it Algebraic geometry,} Springer-Verlag, New
York 1977 (Corollary II.5.18).

\bibitem{HL} Harvey, R., and Lawson Jr., H.B.,
{\it Calibrated geometries,} Acta Math. {\bf 148} (1982), 47--157.

\bibitem{KS} Kashiwara, M., and Schapira, P., {\it Sheaves on manifolds,}
Springer-Verlag, Berlin 1990.

\bibitem{K} Kontsevich, M., {\it Homological algebra of mirror symmetry,}
Proceedings of the 1994 International Congress of Mathematicians {\bf I},
Birkh\"auser, Z\"urich, 1995, p.~120; {\tt alg-} {\tt geom/9411018}.

\bibitem{K2} \bysame talk delivered at ``European Conference on Algebraic
Geometry'', University of Warwick, July 1996.

\bibitem{M} Manin, Yu.I, talk delivered at the Pisa symposium
``Hodge Theory, Mirror Symmetry and Quantum Cohomology'', April 1998.

\bibitem{MDS} McDuff, D., and Salamon, D., {\it Introduction to symplectic
topology,} Clarendon Press, Oxford 1995.

\bibitem{OOY} Ooguri, H., Oz, Y., and Yin, Z., {\it D-branes on Calabi-Yau
spaces and their mirrors,} Nucl.~Phys. {\bf B477} (1996), 407-430; {\tt
hep-th/9606112}.

\bibitem{PZ} Polishchuk, A., and Zaslow, E., {\it Categorical mirror symmetry:
the elliptic curve,} {\tt math.AG/9801119}.

\bibitem{Sad} Sadov, V., {\it On equivalence of Floer's and quantum
cohomology,} Commun.~Math.~Phys. {\bf 173} (1995), 77--99; {\tt
hep-th/9310153.}

\bibitem{SYZ} Strominger, A., Yau, S.-T., and Zaslow, E., {\it  Mirror symmetry
is T-duality,} Nucl.~Phys. {\bf B479} (1996), 243--259; {\tt hep-th/9606040}.

\bibitem{V} Verdier, J.-L., {\it Des cat\'egories d\'eriv\'ees des cat\'egories
ab\'eliennes,} Ast\'erisque {\bf 239}, Soci\'et\'e Math\'ematique de France
(1996).
\end{thebibliography}
\end{document}